%% file: template.tex
\documentclass[cameraready]{Interspeech}

\title{TinyGiantALM: A Compact Audio-Language Model for Intent-Aware Reasoning under Resource Constraints}
\author[affiliation={1,2,3}, orcid=0009-0005-7557-8231]{Vinh-Thuan}{Ly}

\address{
    $^1$ Zalo AI, Vietnam\\
    $^2$ University of Science, VNU-HCM, Ho Chi Minh City, Vietnam \\
    $^3$ Vietnam National University, Ho Chi Minh City, Vietnam 
}

\email{vinhthuanly210@gmail.com}

\usepackage{comment}

\begin{document}

\maketitle

\input{Section/00_Abtract}

\input{Section/01_Intro}

\input{Section/02_Relatedwork}

\input{Section/03_Method}
\input{Section/04_Experiment}
\input{Section/05_Conclusion}
\input{Section/06_GenAIUsage}

\bibliographystyle{IEEEtran}
\bibliography{mybib}

\end{document}

%% file: Section/00_Abtract.tex
\begin{abstract}
Current advancements in Audio Reasoning rely on massive Large Audio-Language Models (LALMs), hindering deployment in resource-constrained environments. We introduce TinyGiantALM, a compact 1.5B efficiency-oriented alternative. Instead of brute-force scaling, we propose an Instruction-Aware Feature Refinement framework using a Query-guided Projector and Semantic Gating to filter acoustic signals based on user intent. On the MMAR benchmark, TinyGiantALM achieves 46.4\% zero-shot accuracy, significantly outperforming 7B--13B baselines. While a reasoning gap in logical narrative remains versus 30B+ models and certain trade-offs exist in overly dense or spatial scenes, our approach notably surpasses models up to $8\times$ larger in disentangling mixed-modality environments. These findings demonstrate that architectural precision offers a tangible pathway to secure robust perception capabilities on edge-friendly scales.
\end{abstract}

%% file: Section/01_Intro.tex
\section{Introduction}

The recent \textit{Interspeech 2026 Audio Reasoning Challenge}\cite{ma2026interspeech2026audioreasoning} highlighted a pivotal shift in auditory intelligence, where Large Audio-Language Models (LALMs) like Qwen3-Omni achieve remarkable reasoning depth. However, these top-tier systems predominantly rely on massive parameter scaling ($>$7B--30B) and computationally expensive Reinforcement Learning, creating significant barriers for deployment in resource-constrained environments. This trend prompts a critical question: \textit{Can architectural priors compensate for reduced scale to achieve viable audio reasoning on edge devices?}


To address this, we propose \textbf{TinyGiantALM}, a compact 1.5B model designed to explore the efficiency-reasoning trade-off. Unlike challenge winners that leverage brute-force scaling, our approach focuses on \textit{Instruction-Aware Feature Refinement}. We hypothesize that for smaller models, the key to reasoning is not vast memorization, but the active filtering of acoustic noise based on user intent.

Concretely, our architecture combines a \textbf{Query-guided Projector} with a \textbf{CLAP-driven Semantic Gating} mechanism. Rather than passively processing all audio tokens, the model conditions acoustic features on the textual query, so that the Small Language Model (SLM) attends only to task-relevant signals.

Empirically, while a performance gap remains compared to massive challenge leaders (46.40\% vs. 74.00\% on MMAR \cite{ma2025mmarchallengingbenchmarkdeep}), TinyGiantALM effectively outperforms traditional baselines of significantly larger size (e.g., SALMONN-13B) in mixed-modality tasks. Our contributions are:

\begin{itemize}[leftmargin=*]
    \item We present \textbf{TinyGiantALM} (1.5B), offering a study on the trade-offs between parameter scale and reasoning depth within the Challenge context.
    \item We propose a \textbf{Semantic Gating mechanism} that aligns acoustic features with user intent, reducing the "blindness" of compact models in complex scenes.
    \item We demonstrate that specialized architectures can secure robust perception capabilities on edge-friendly scales, providing a lightweight alternative to server-grade LALMs.
\end{itemize}

\begin{figure}[t]
    \centering
    \includegraphics[width=1\linewidth]{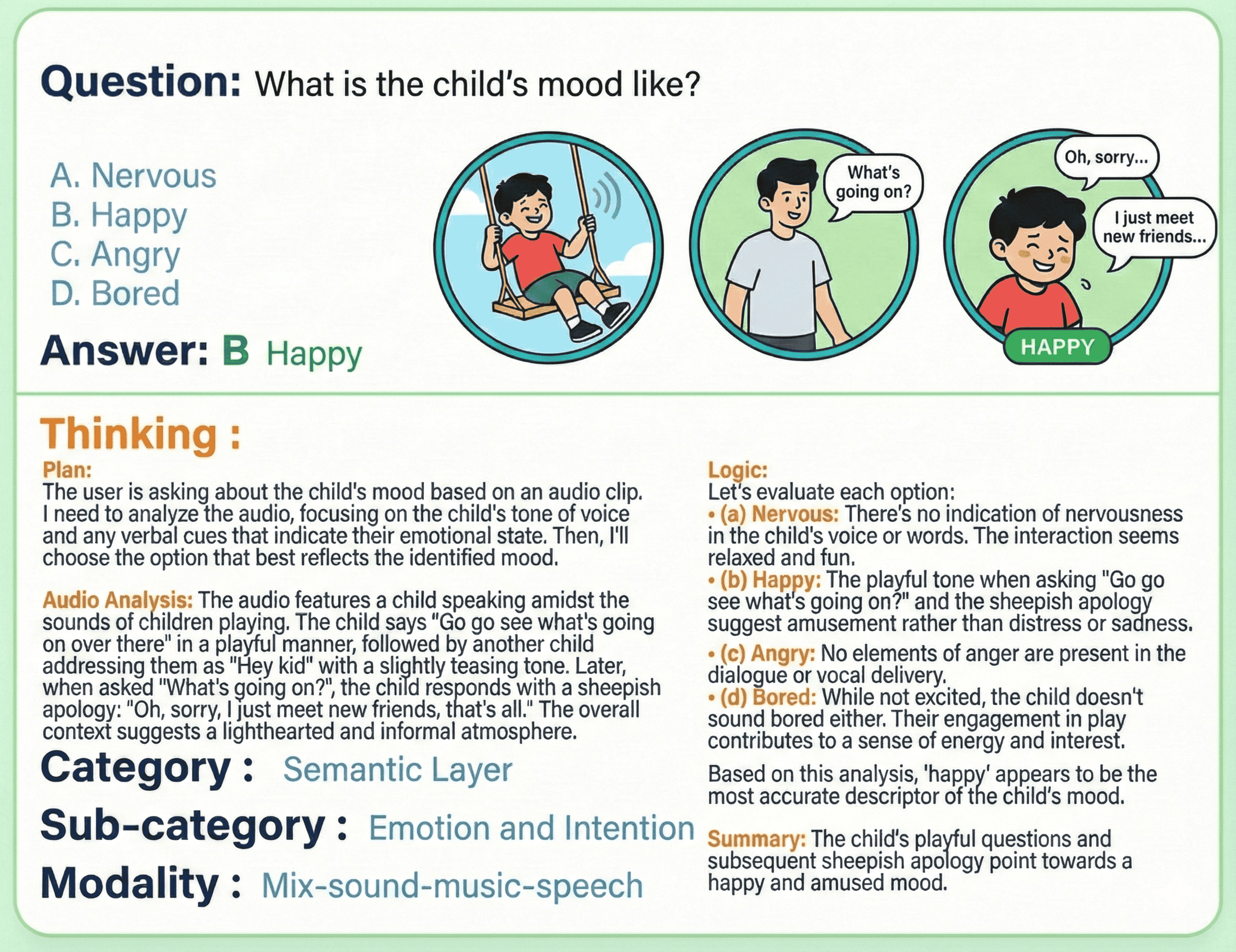}
    \vspace{-20pt}
    \caption{A sample output from TinyGiantALM.}
    \label{fig:intro}
    \vspace{-20pt}
\end{figure}

%% file: Section/02_Relatedwork.tex
\section{Related Work}

\noindent\textbf{Deep Reasoning in LALMs.} 
The success of Chain-of-Thought (CoT) \cite{NEURIPS2022_9d560961} has shifted audio research from simple perception toward cognitive reasoning. Despite this, audio remains challenging due to high temporal density and overlapping signals. Current state-of-the-art LALMs, such as SALMONN \cite{tang2024salmonngenerichearingabilities} and Qwen2-Audio \cite{chu2024qwen2audiotechnicalreport}, rely on massive parameter scaling (7B-13B) to bridge the semantic gap. However, as shown by the MMAR benchmark \cite{ma2025mmarchallengingbenchmarkdeep}, these brute-force approaches often falter in complex scenes, struggling to disentangle acoustic events without explicit reasoning. Our work diverges by prioritizing architectural efficiency over parameter scale.

\noindent\textbf{Efficient Temporal Modeling.} 
Effective audio reasoning requires capturing both transient acoustic details and long-term semantic dependencies. While architectures like Conformer \cite{gulati2020conformerconvolutionaugmentedtransformerspeech} and E-Branchformer \cite{10022656} have set benchmarks in speech processing by decoupling local-global contexts, their potential as "reasoning bridges" for LLMs remains largely unexplored. Most existing ALMs utilize simple linear projectors that fail to filter task-relevant information. TinyGiantALM addresses this by repurposing the E-Branchformer as an instruction-aware projector, integrating it with a semantic gating mechanism to achieve deep reasoning within a compact 1.5B footprint.

%% file: Section/03_Method.tex
\section{Methodology}
\label{sec:method}

\begin{figure*}[t]
    \centering
    \includegraphics[width=1.\linewidth]{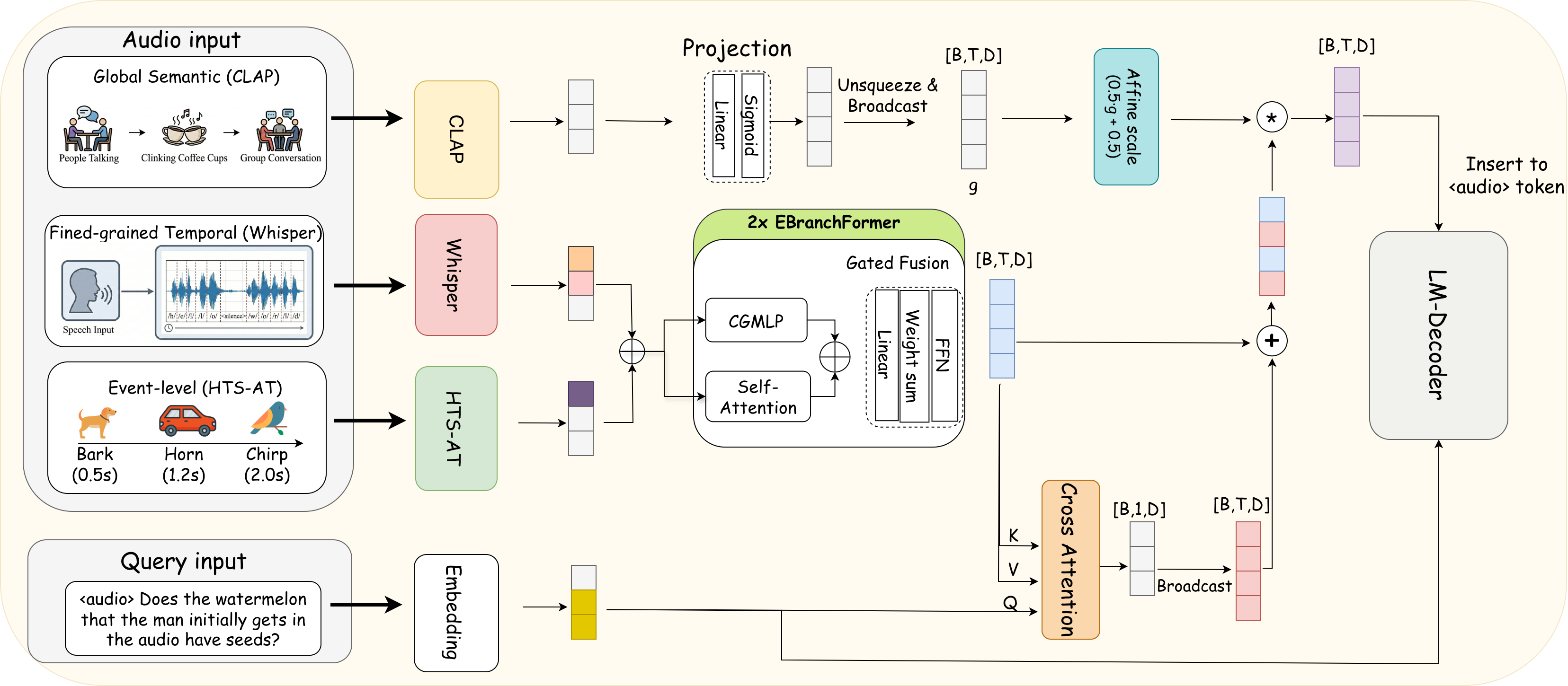} 
    \vspace{-15pt}
    \caption{Proposed TinyGiantALM architecture. (1) Triple-stream frontend (Whisper, HTS-AT, CLAP) extracts multi-scale features. (2) E-Branchformer blocks encode local-global contexts via CGMLP and Self-Attention. (3) A User Intent query drives Cross-Attention, while a CLAP-based gate ($\mathbf{g}$) modulates features via affine scaling ($0.5 \cdot \mathbf{g} + 0.5$) before LLM injection for CoT reasoning.}
    \vspace{-15pt}
    \label{fig:architecture}
\end{figure*}

We propose TinyGiantALM, an audio-language model designed to bridge the gap between heterogeneous acoustic signals and textual user intent. As illustrated in Figure \ref{fig:architecture}, our framework consists of a triple-stream acoustic front-end, a novel Query-guided Projector utilizing E-Branchformer blocks, and a Large Language Model (LLM) backbone.

\subsection{Triple-stream Acoustic Feature Extraction}
\label{subsec:features}
To capture the full spectrum of auditory information—from fine-grained phonemes to holistic environmental context—we employ a multi-rate resampling strategy integrating three frozen pre-trained encoders (totaling \textbf{732M} parameters).

\noindent\textbf{Fine-grained Temporal Stream ($\mathbf{F}_w$).} 
We utilize the encoder of \textit{Whisper-Large-v3-turbo} \cite{pmlr-v202-radford23a} to extract linguistic and paralinguistic details. The input audio is resampled to 16kHz, and the resulting representation captures detailed phonetic content essential for speech-heavy reasoning tasks:
\begin{equation}
\setlength{\abovedisplayskip}{3pt}
\setlength{\belowdisplayskip}{3pt}
    \mathbf{F}_w = \operatorname{Encoder}_{\text{Whisper}}(\mathbf{X}_{16k}) \in \mathbb{R}^{T_w \times 1280}
\end{equation}

\noindent\textbf{Event-level Stream ($\mathbf{F}_h$).} 
To perceive short-duration acoustic events (e.g., sirens, bird chirps) often missed by speech models, we employ \textit{HTS-AT} \cite{9746312}, a hierarchical transformer trained on AudioSet. Processing 48kHz audio, it provides event-aware features with high temporal sensitivity:
\begin{equation}
\setlength{\abovedisplayskip}{3pt}
\setlength{\belowdisplayskip}{3pt}
    \mathbf{F}_h = \operatorname{Encoder}_{\text{HTS-AT}}(\mathbf{X}_{48k}) \in \mathbb{R}^{T_h \times 768}
\end{equation}

\noindent\textbf{Global Semantic Stream ($\mathbf{c}_{clap}$).} 
Unlike temporal streams, the \textit{CLAP} encoder \cite{10095889} condenses audio into a single semantic anchor. This holistic summary (e.g., "speech in noise") acts as a global prior for our gating mechanism:
\begin{equation}
\setlength{\abovedisplayskip}{3pt}
\setlength{\belowdisplayskip}{3pt}
    \mathbf{c}_{clap} = \operatorname{Encoder}_{\text{CLAP}}(\mathbf{X}_{48k}) \in \mathbb{R}^{1024}
\end{equation}

To ensure alignment across heterogeneous streams, we apply \textit{Adaptive Average Pooling} to fix the sequence length of both temporal streams to $N=300$ tokens before concatenation.
\subsection{Query-guided Triple-stream Projector}
\label{subsec:projector}
Our projector $\mathcal{P}_\theta$ maps acoustic features to the LLM space $\mathbb{R}^{d_{model}}$ via three dynamic stages.

\noindent\textbf{Stage A: E-Branchformer Encoding.}
We project the concatenated temporal streams to the LLM latent dimension $d_{model} = 1024$:
\begin{equation}
    \mathbf{H}_{0} = \operatorname{Linear}([\mathbf{F}_w; \mathbf{F}_h]) \in \mathbb{R}^{T \times d_{model}}
\end{equation}
We employ $L=2$ E-Branchformer blocks \cite{10022656} to model local-global dependencies. For an input $\mathbf{X}$, the block utilizes a dual-path architecture:

\textit{1) Global Branch (MHSA):} Captures long-range context via Multi-Head Self-Attention:
\begin{equation}
    \mathbf{Z}_{attn} = \operatorname{LayerNorm}(\mathbf{X} + \operatorname{Dropout}(\operatorname{MHSA}(\mathbf{X})))
\end{equation}

\textit{2) Local Branch (Convolution):} Captures acoustic transients via 1D depth-wise convolution ($k=17$). Features are expanded, convolved, and projected back:
\begin{equation}
    \mathbf{X}_{conv} = \operatorname{Lin}_{2}(\sigma(\operatorname{DConv}_{17}(\sigma(\operatorname{Lin}_{1}(\mathbf{X})))))
\end{equation}
\begin{equation}
    \mathbf{Z}_{conv} = \operatorname{LayerNorm}(\mathbf{X} + \operatorname{Dropout}(\mathbf{X}_{conv}))
\end{equation}

\textit{3) Dynamic Merge:} Branch outputs are dynamically fused using weights derived from their concatenated features:
\begin{equation}
    [\mathbf{W}_{attn}, \mathbf{W}_{conv}] = \operatorname{softmax}(\operatorname{Linear}([\mathbf{Z}_{attn}; \mathbf{Z}_{conv}]))
\end{equation}
\begin{equation}
    \mathbf{Z}_{merge} = \mathbf{W}_{attn} \odot \mathbf{Z}_{attn} + \mathbf{W}_{conv} \odot \mathbf{Z}_{conv}
\end{equation}

\textit{4) Final FFN:} The fused representation passes through a Feed-Forward Network ($\operatorname{FFN}$):
\begin{equation}
    \mathbf{H}_{enc} = \operatorname{LayerNorm}(\mathbf{Z}_{merge} + \operatorname{Dropout}(\operatorname{FFN}(\mathbf{Z}_{merge})))
\end{equation}
This hybrid encoding robustly aligns the acoustic scene and temporal events into the LLM space.

\noindent\textbf{Stage B: User Intent-Aware Refinement.} 
To align the audio representation with the user's specific instruction, we derive a global User Intent vector $\mathbf{q}_{intent}$ by applying masked mean pooling over the user's text instruction tokens ($M_{user}$):
\begin{equation}
    \mathbf{q}_{intent} = \frac{1}{|M_{user}|} \sum_{i \in M_{user}} \mathbf{E}_{text}^{(i)}
\end{equation}
We then employ a Multi-Head Cross-Attention mechanism. The User Intent $\mathbf{q}_{intent}$ acts as the Query, while the encoded audio $\mathbf{H}_{enc}$ serves as both Key and Value. This acts as a residual refinement to highlight semantically relevant segments:
\begin{equation}
    \mathbf{Attn} = \operatorname{CrossAttention}(\mathbf{q}_{intent}, \mathbf{H}_{enc}, \mathbf{H}_{enc})
\end{equation}
\begin{equation}
    \mathbf{H}_{query} = \operatorname{LayerNorm}(\mathbf{H}_{enc} + \mathbf{Attn})
\end{equation}

\noindent\textbf{Stage C: CLAP-driven Semantic Gating.} 
Finally, to ground features in the global context, we utilize the CLAP anchor. We compute a soft gate $\mathbf{g} \in (0, 1)$ via a sigmoid projection:
\begin{equation}
    \mathbf{g} = \sigma(\operatorname{Linear}(\mathbf{c}_{clap}))
\end{equation}
The features are modulated via affine scaling to preserve signal integrity while injecting global context:
\begin{equation}
    \mathbf{H}_{final} = \operatorname{LayerNorm}\left(\mathbf{H}_{query} \odot (0.5 + 0.5 \cdot \mathbf{g})\right)
\end{equation}

\subsection{LLM Integration and Training Strategy}
\label{subsec:llm}
The refined embeddings $\mathbf{H}_{final}$ are inserted at the \texttt{<audio>} token position into the \textbf{Qwen3 (0.6B)} backbone. We train the model to adhere to the CoT format of the CoTA dataset \cite{zhifei-etal-2025-audio}, encapsulating reasoning steps within \texttt{Plan},\texttt{ Audio Analysis}, \texttt{Logic}, and \texttt{Summary} tags inside a \texttt{<think>} block. The model is optimized via Next Token Prediction on the assistant's response, masking user instructions during loss computation.

%% file: Section/04_Experiment.tex
\section{Experiment}
\subsection{Dataset}
\label{subsec:dataset}


\noindent\textbf{Training Data.} We utilize all publicly available subsets of the CoTA dataset \cite{zhifei-etal-2025-audio} hosted on Hugging Face, totaling \textbf{558,423} instruction-tuning samples. This collection excludes AudioSet and comprises: (1) \textit{Environmental Sounds} (AudioCaps \cite{kim-etal-2019-audiocaps}, Clotho \cite{9052990}); (2) \textit{Speech} (MELD \cite{poria-etal-2019-meld}, CoVoST 2 \cite{wang2020covost2massivelymultilingual}); and (3) \textit{Music} (MusicBench \cite{melechovsky-etal-2024-mustango}). The data follows a structured pipeline modeling \textit{Planning}, \textit{Captioning}, \textit{Reasoning}, and \textit{Summary} phases to foster deep logical inference.


\subsection{Implementation Details}
Experiments are conducted on a single NVIDIA A100 GPU using Qwen3-0.6B as the backbone, fine-tuned for 3 epochs. During inference, the model requires only ~5GB VRAM, confirming its edge-friendly memory footprint. We employ the AdamW optimizer with decoupled learning rates ($1 \times 10^{-4}$ for the projector, $5 \times 10^{-5}$ for the LLM) and enable BFloat16 precision with TF32. The effective batch size is set to 32, with maximum sequence lengths of 300 audio frames and 2048 text tokens.

\subsection{Main Results}
\label{subsec:results}


We evaluate TinyGiantALM on the challenging MMAR benchmark. Alongside our official Challenge standing (Table \ref{tab:challenge_leaderboard}), Table \ref{tab:main_results} summarizes our zero-shot performance against state-of-the-art baselines. Evaluating our architecture across these paradigms reveals three critical advantages:

\begin{table}[ht] 
    \vspace{-8pt}
    \centering
    \caption{Zero-shot accuracy (\%) on MMAR. S: Sound, M: Music, Sp: Speech. TinyGiantALM is \textbf{1.5B}. Best open-source performance is \textbf{bolded}, second best is \underline{underlined}.}
    \vspace{-8pt}
    \resizebox{\linewidth}{!}{
    \begin{tabular}{l|c|ccc|cccc|c}
    \toprule
    \textbf{Model} & \textbf{Sz} & \multicolumn{3}{c|}{\textbf{Single}} & \multicolumn{4}{c|}{\textbf{Mixed}} & \textbf{Avg} \\
    & & S & M & Sp & S-M & S-Sp & M-Sp & All & \\
    \midrule
    \multicolumn{10}{l}{\textbf{(a) Large Audio-Language Models (LALMs)}} \\
    \midrule
    Flamingo 2 \cite{ghosh2025audioflamingo2audiolanguage} & 3B & 24.9 & 17.5 & 20.8 & 18.2 & 26.6 & 23.2 & 8.3 & 21.9 \\
    LTU-AS \cite{10389742} & 7B & 20.0 & 14.1 & 19.1 & 9.1 & 20.6 & 28.1 & 12.5 & 19.0 \\
    GAMA \cite{ghosh2024gamalargeaudiolanguagemodel} & 7B & 29.1 & 24.3 & 27.9 & 27.3 & 24.8 & 28.1 & 20.8 & 26.5 \\
    Qwen2-Audio \cite{chu2024qwen2audiotechnicalreport} & 8.4B& 33.3 & 24.3 & 32.3 & 9.1 & 31.2 & 30.5 & 25.0 & 30.0 \\
    SALMONN \cite{tang2024salmonngenerichearingabilities} & 13B& 30.3 & 31.1 & 34.7 & 9.1 & 34.9 & 35.4 & \underline{41.7} & 33.2 \\
    GPT-4o mini Audio \cite{10.1007/978-3-031-92611-2_4} & - & 38.8 & 35.9 & 58.8 & 45.5 & 60.1 & 57.3 & 50.0 & 50.6 \\
    \midrule
    \multicolumn{10}{l}{\textbf{(b) Large Audio Reasoning Models (LARMs)}} \\
    \midrule
    Audio-CoT \cite{ma2025audiocotexploringchainofthoughtreasoning} & 8.4B& 35.8 & 25.2 & 34.0 & 9.1 & 30.7 & 30.5 & 37.5 & 31.3 \\
    Audio-Reasoner \cite{zhifei-etal-2025-audio} & 8.4B& 43.6 & 33.5 & 33.0 & \underline{45.5} & 42.7 & 31.7 & 25.0 & 36.8 \\
    \midrule
    \multicolumn{10}{l}{\textbf{(c) Omni \& Large Language Models (OLMs)}} \\
    \midrule
    Baichuan-Omni-1.5 \cite{li2025baichuanomni15technicalreport} & 11B& 41.2 & 33.0 & 40.5 & 36.4 & 48.6 & 39.0 & \underline{41.7} & 40.7 \\
    Cap+DeepSeek-V3 \cite{deepseekai2025deepseekv3technicalreport} & 671B& 42.4 & \textbf{40.8} & \underline{56.1} & 18.2 & \underline{50.0} & 45.1 & 37.5 & \underline{47.6} \\
    Qwen2.5-Omni \cite{xu2025qwen25omnitechnicalreport} & 7B & \textbf{58.8} & \textbf{40.8} & \textbf{59.9} & \textbf{54.6} & \textbf{61.9} & \textbf{67.1} & \textbf{58.3} & \textbf{56.7} \\
    Gemini 2.0 Flash \cite{geminiteam2025geminifamilyhighlycapable} & -  & 61.2 & 51.0 & 72.1 & 81.8 & 72.5 & 65.9 & 70.8 & 65.6 \\
    \midrule\midrule
    \textbf{TinyGiantALM (Ours)} & \textbf{1.5B}& \underline{47.3} & \underline{37.9} & 46.9 & \underline{45.5} & 49.1 & \underline{58.5} & \underline{41.7} & 46.4 \\
    \bottomrule
    \end{tabular}
    }
    \label{tab:main_results}
    \vspace{-10pt}
\end{table}

\begin{figure}[t]
    \centering
    \includegraphics[width=1\linewidth]{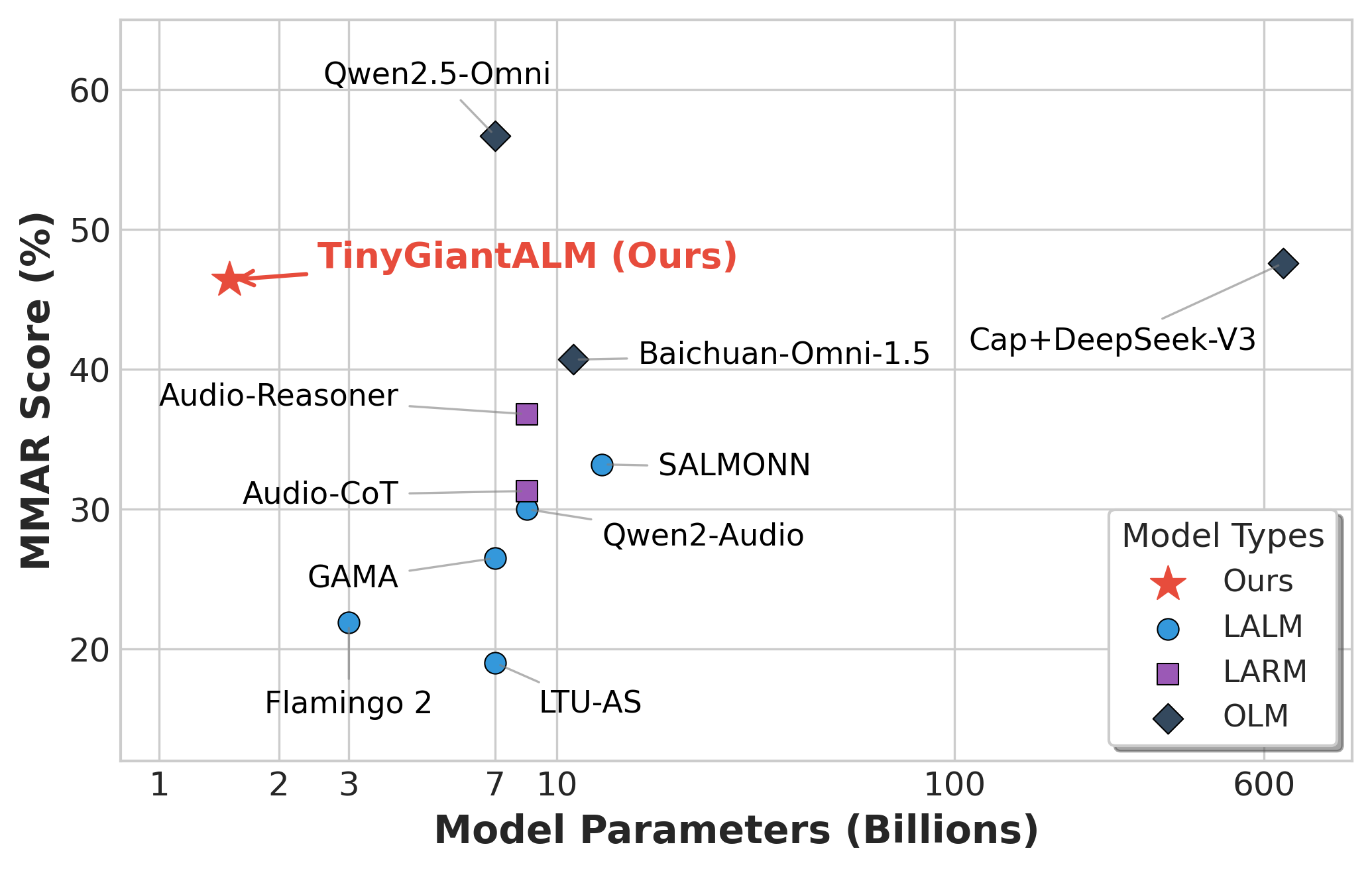}
    \vspace{-20pt}
    \caption{Zero-shot MMAR Accuracy vs. Model Size.}
    \label{fig:tradeoff}
    \vspace{-20pt}
\end{figure}

\noindent\textbf{Efficiency vs. LALMs:} Traditional LALMs exhibit a severe performance collapse in complex scenarios. Notably, Qwen2-Audio \cite{chu2024qwen2audiotechnicalreport} (8.4B) and SALMONN \cite{tang2024salmonngenerichearingabilities} (13B) drop to \textbf{9.1\%} accuracy in \textit{Mix-Sound-Music} tasks, indicating a failure to disentangle overlapping streams. In contrast, TinyGiantALM (1.5B) maintains a robust \textbf{45.5\%}, outperforming these baselines by over \textbf{+36\%} despite being $5\times$--$8\times$ smaller. Overall, as illustrated in Figure~\ref{fig:tradeoff}, our model achieves \textbf{46.4\%}, setting a new efficiency standard.
    
\noindent\textbf{Superior Reasoning vs. LARMs:} Compared to specialized Audio Reasoning models, TinyGiantALM demonstrates superior logical deduction. It surpasses Audio-Reasoner (8.4B, 36.8\%) by a significant margin of \textbf{+9.6\%}. This confirms that our \textit{Query-guided Triple-stream Projector} provides richer, more disentangled cues for reasoning than standard single-stream encoders.
    
\noindent\textbf{Closing the Gap with OLMs:} Remarkably, our 1.5B model rivals massive Omni-models. It outperforms Baichuan-Omni-1.5 \cite{li2025baichuanomni15technicalreport} (11B, 40.7\%) and approaches the performance of the 671B Caption+DeepSeek-V3 \cite{deepseekai2025deepseekv3technicalreport} pipeline (47.6\%). Specifically in the \textit{Mix-Music-Speech} domain, TinyGiantALM achieves \textbf{58.5\%}, trailing only the SOTA Qwen2.5-Omni \cite{xu2025qwen25omnitechnicalreport}, proving that specialized architectural priors can compensate for the lack of massive pre-training scale (see Figure~\ref{fig:tradeoff}).

\begin{table}[ht]
    \vspace{-8pt}
    \centering
    \caption{Comparison with Top-Tier Challenge Solutions (Single Model Track)\cite{ma2026interspeech2026audioreasoning}. \textbf{Size (Est.)} indicates parameter count. \textbf{Rubrics} denotes reasoning quality, \textbf{Acc} denotes final accuracy.}
    \vspace{-8pt}
    \resizebox{\linewidth}{!}{
    \begin{tabular}{c|l|c|cc}
    \toprule
    \textbf{Rank} & \textbf{Backbone / Method} & \textbf{Size (Est.)} & \textbf{Rubrics} & \textbf{Acc} \\
    \midrule
    1 & Qwen3-Omni\cite{xu2025qwen3omnitechnicalreport} + RL (GRPO) & ($>$30B) & 65.29 & 74.00 \\
    2 & Qwen3-Omni\cite{xu2025qwen3omnitechnicalreport} + Attn. Manip. & ($>$30B) & 62.55 & 71.00 \\
    3 & Qwen3-Omni\cite{xu2025qwen3omnitechnicalreport} + LoRA & ($>$30B) & 62.22 & 71.70 \\
    \midrule
    13 & \textbf{TinyGiantALM (Ours)} & \textbf{1.5B} & \textbf{23.77} & \textbf{46.40} \\
    \bottomrule
    \end{tabular}
    }
    \label{tab:challenge_leaderboard}
    \vspace{-20pt}
\end{table}



\noindent\textbf{Reasoning Gap vs. Scale.}
As shown in Table~\ref{tab:challenge_leaderboard}, top teams leverage massive \textit{Qwen3-Omni}\cite{xu2025qwen3omnitechnicalreport} backbones ($>$30B) and costly RL strategies to master the explicit, multi-step evidence tracking required for high reasoning scores (Rubrics $>$62.00\%). In transparent comparison, our 1.5B model's lower Rubrics score (23.77\%, Rank 13) highlights a clear intrinsic limitation of compact architectures: while they can perceive complex environments, they struggle to generate and maintain the long-context, highly detailed logical narratives that foundation models (scaled orders of magnitude larger) can easily produce.

\noindent\textbf{Efficiency-Oriented Competitiveness.}
Despite this expected reasoning gap, TinyGiantALM offers highly competitive viability for edge scenarios. Operating with only \textbf{1.5B parameters}, which is \textbf{approximately $\mathbf{1/20^{th}}$ the size} of the leading challenge baselines in Table~\ref{tab:challenge_leaderboard}, it remarkably retains \textbf{62.7\%} of the State-of-the-Art accuracy (46.40\% vs. 74.00\%). This performance is particularly striking given that many traditional mid-scale LALMs (7B--13B) collapse in mixed-modality tasks. This confirms that while massive parameter scale is undeniably necessary for exhaustive deep reasoning, strategically designed architectural priors can effectively secure robust, lightweight perception capabilities in resource-constrained environments.

\noindent\textbf{Analyzing the Rubric-Accuracy Gap.}
The notable divergence between the final task accuracy (46.40\%) and the intermediate reasoning score (23.77\%) warrants deeper analysis. The gap suggests that while 1.5B models can identify correct causal relations (Accuracy), they struggle with linguistic richness in reasoning chains (Rubrics) due to limited language modeling capacity. As illustrated in Figure~\ref{fig:intro}, TinyGiantALM correctly infers a "Happy" mood by synthesizing high-level semantics (e.g., recognizing a "playful tone") and successfully applying elimination logic. However, by jumping directly to holistic, goal-oriented conclusions, the model often omits the granular, low-level acoustic triggers (such as explicitly naming "laughter" or specific background noises) that the comprehensive MMAR-Rubrics strictly require and reward. Consequently, while TinyGiantALM lacks the descriptive verbosity and exhaustive narration of massive 30B+ models, its concise, direct reasoning remains highly pragmatic and sufficiently accurate for practical edge applications.

\subsection{Ablation Study}
\label{subsec:ablation}

We isolate the contributions of the \textbf{Inference Query (IQ)} and \textbf{CLAP-based Global Gate} via ablation in Table~\ref{tab:ablation_final}.

\begin{table}[ht]
    \vspace{-8pt}
    \centering
    \caption{Ablation study on MMAR Modalities. Columns are sorted by ascending overall performance. \textbf{Vanilla}: No IQ \& No CLAP. \textbf{w/o IQ}: Has CLAP, no Query. \textbf{w/o CLAP}: Has Query, no CLAP. \textbf{$\Delta$}: Full vs Vanilla.}
    \vspace{-8pt}
    \resizebox{\linewidth}{!}{
    \begin{tabular}{l|cccc|c}
    \toprule
    \textbf{Modality} & \textbf{Vanilla} & \textbf{w/o IQ} & \textbf{w/o CLAP} & \textbf{Full Model} & \textbf{$\Delta$} \\
    \midrule
    \textit{Single Modality:} & & & & & \\
    Sound (S) & 33.94 & 38.79 & 38.79 & \textbf{47.27} & +13.33 \\
    Music (M) & 35.92 & 33.50 & 35.44 & \textbf{37.86} & +1.94 \\
    Speech (Sp) & 39.80 & 44.56 & 43.54 & \textbf{46.94} & +7.14 \\
    \midrule
    \textit{Mixed Modality:} & & & & & \\
    Mix S-M & 9.09 & 18.18 & 36.36 & \textbf{45.45} & \textbf{+36.36} \\
    Mix S-Sp & 38.99 & 40.37 & 47.71 & \textbf{49.08} & +10.09 \\
    Mix M-Sp & 43.90 & 36.59 & 50.00 & \textbf{58.54} & +14.64 \\
    Mix All & 45.83 & 41.67 & \textbf{54.17} & 41.67 & -4.16 \\
    \midrule
    \textbf{Overall Accuracy} & \textbf{38.00} & \textbf{39.70} & \textbf{42.70} & \textbf{46.40} & \textbf{+8.40} \\
    \bottomrule
    \end{tabular}
    }
    \label{tab:ablation_final}
    \vspace{-10pt}
\end{table}

\begin{table}[ht]
    \vspace{-8pt}
    \centering
    \caption{Detailed performance breakdown across all 16 MMAR sub-tasks, sorted by model complexity. \textbf{$\Delta$} shows the improvement of Full Model over Vanilla.}
    \vspace{-8pt}
    \resizebox{\linewidth}{!}{
    \begin{tabular}{l|cccc|c}
    \toprule
    \textbf{Sub-Task} & \textbf{Vanilla} & \textbf{w/o IQ} & \textbf{w/o CLAP} & \textbf{Full Model} & \textbf{$\Delta$} \\
    \midrule
    \multicolumn{6}{l}{\textit{\textbf{1. Signal Layer Tasks}}} \\
    Acoustic Quality Analysis & 27.78 & 22.22 & \textbf{38.89} & 33.33 & +5.55 \\
    Audio Difference Analysis & 25.00 & 25.00 & 50.00 & \textbf{62.50} & \textbf{+37.50} \\
    Spatial Analysis & \textbf{40.00} & 26.67 & \textbf{46.67} & 33.33 & -6.67 \\
    Temporal Analysis & 42.86 & 42.86 & 35.71 & \textbf{46.43} & +3.57 \\
    \midrule
    \multicolumn{6}{l}{\textit{\textbf{2. Perception Layer Tasks}}} \\
    Anomaly Detection & \textbf{41.18} & 23.53 & \textbf{41.18} & \textbf{41.18} & 0.00 \\
    Counting & 29.29 & 26.26 & \textbf{40.40} & 36.36 & +7.07 \\
    Env. Perception & 31.54 & 39.60 & \textbf{44.30} & 43.62 & +12.08 \\
    Speaker Analysis & 39.58 & 41.67 & 41.67 & \textbf{45.83} & +6.25 \\
    \midrule
    \multicolumn{6}{l}{\textit{\textbf{3. Semantic Layer Tasks}}} \\
    Content Analysis & 45.07 & 45.39 & 49.34 & \textbf{51.97} & +6.90 \\
    Correlation Analysis & \textbf{52.00} & 30.00 & 38.00 & 50.00 & -2.00 \\
    Emotion \& Intention & 40.00 & 46.67 & 38.33 & \textbf{51.67} & +11.67 \\
    Professional Knowledge & 33.80 & 38.03 & 42.25 & \textbf{46.48} & +12.68 \\
    \midrule
    \multicolumn{6}{l}{\textit{\textbf{4. Cultural Layer Tasks}}} \\
    Aesthetic Evaluation & 37.50 & 50.00 & 50.00 & \textbf{62.50} & \textbf{+25.00} \\
    Culture of Speaker & 42.31 & 32.69 & 44.23 & \textbf{57.69} & +15.38 \\
    Imagination & \textbf{30.00} & \textbf{30.00} & \textbf{30.00} & \textbf{30.00} & 0.00 \\
    Music Theory & 22.22 & 15.87 & 22.22 & \textbf{31.75} & +9.53 \\
    \midrule
    \textbf{Average Accuracy} & \textbf{38.00} & \textbf{39.70} & \textbf{42.70} & \textbf{46.40} & \textbf{+8.40} \\
    \bottomrule
    \end{tabular}
    }
    \label{tab:full_breakdown}
    \vspace{-15pt}
\end{table}

\noindent\textbf{Synergy \& Modality Disentanglement.} 
As detailed in Table~\ref{tab:ablation_final}, adding IQ or the CLAP gate individually yields moderate gains over the Vanilla baseline (38.00\%), but their combination unlocks a non-linear boost to \textbf{46.40\%} ($\Delta$+8.40\%). This suggests IQ primes the attention layers, enabling CLAP's semantic anchor for precise deduction. Crucially, it resolves the "Cocktail Party Problem" in mixed environments. For instance, in \textit{Mix S-M}, the Vanilla model collapses to near-random guessing (9.09\%), while our Full Model restores performance to \textbf{45.45\%} ($\Delta$+36.36\%), proving its ability to effectively disentangle overlapping streams.

\noindent\textbf{Advancement in Higher-Order Cognition.} 
Table~\ref{tab:full_breakdown} reveals substantial gains in multi-step reasoning tasks. In \textit{Audio Difference Analysis}, the model improves by \textbf{+37.50\%}, indicating IQ guides attention to specific temporal changes. \textit{Aesthetic Evaluation} sees a surge of \textbf{+25.00\%}, as the CLAP gate introduces rich, text-aligned priors about artistic quality. Furthermore, a steady improvement in \textit{Professional Knowledge} (+12.68\%) suggests that grounding generation in global semantics significantly reduces hallucinations for domain-specific terminology.



\noindent\textbf{Gating Limitations.} Gating entails trade-offs: in "Mix All," the Full Model drops 4.16\% vs. Vanilla (w/o CLAP: 54.17\%), as global CLAP anchors become noisy in dense scenes. Aggressive filtering also strips fine physical cues, causing regressions in \textit{Spatial Analysis} (-6.67\%) and \textit{Correlation} (-2.00\%).

%% file: Section/05_Conclusion.tex
\section{Conclusion}
\label{sec:conclusion}

In this work, we presented TinyGiantALM, a 1.5B model exploring the trade-offs between efficiency and reasoning depth. While our architecture achieves a competitive 46.40\% zero-shot MMAR accuracy, outperforming significantly larger baselines, we acknowledge a persistent reasoning gap reflected in our 23.77\% Rubric score compared to 30B+ foundation models. This suggests that while specialized priors can effectively secure robust perception, the generation of exhaustive, multi-step logical narratives remains constrained by language modeling scale. Furthermore, we identify inherent limitations of semantic gating in overly dense "Mix All" environments (-4.16\%) and spatial tasks, where global anchors may introduce noise. Although current evaluations rely on server-grade hardware, the compact 1.5B footprint establishes a baseline for future on-device optimization. We hope this study encourages intent-aware architectures that democratize auditory intelligence beyond brute-force scaling.

%% file: Section/06_GenAIUsage.tex
\section{Use of Generative AI Disclosure}
Generative AI tools were used for linguistic polishing and preparing illustrative elements. All scientific content, experimental design, and results were produced by the author.

%% file: mybib.bib
@misc{deepseekai2025deepseekv3technicalreport,
      title={DeepSeek-V3 Technical Report}, 
      author={DeepSeek-AI and Aixin Liu and Bei Feng and Bing Xue and others},
      year={2025},
      eprint={2412.19437},
      archivePrefix={arXiv},
      primaryClass={cs.CL},
      url={https://arxiv.org/abs/2412.19437}, 
}

@misc{tang2024salmonngenerichearingabilities,
      title={SALMONN: Towards Generic Hearing Abilities for Large Language Models}, 
      author={Changli Tang and Wenyi Yu and Guangzhi Sun and Xianzhao Chen and Tian Tan and Wei Li and Lu Lu and Zejun Ma and Chao Zhang},
      year={2024},
      eprint={2310.13289},
      archivePrefix={arXiv},
      primaryClass={cs.SD},
      url={https://arxiv.org/abs/2310.13289}, 
}

@misc{chu2024qwen2audiotechnicalreport,
      title={Qwen2-Audio Technical Report}, 
      author={Yunfei Chu and Jin Xu and Qian Yang and Haojie Wei and Xipin Wei and Zhifang Guo and Yichong Leng and Yuanjun Lv and Jinzheng He and Junyang Lin and Chang Zhou and Jingren Zhou},
      year={2024},
      eprint={2407.10759},
      archivePrefix={arXiv},
      primaryClass={eess.AS},
      url={https://arxiv.org/abs/2407.10759}, 
}

@misc{ma2025mmarchallengingbenchmarkdeep,
      title={MMAR: A Challenging Benchmark for Deep Reasoning in Speech, Audio, Music, and Their Mix}, 
      author={Ziyang Ma and Yinghao Ma and Yanqiao Zhu and others},
      year={2025},
      eprint={2505.13032},
      archivePrefix={arXiv},
      primaryClass={cs.SD},
      url={https://arxiv.org/abs/2505.13032}, 
}

@INPROCEEDINGS{10022656,
  author={Kim, Kwangyoun and Wu, Felix and Peng, Yifan and Pan, Jing and Sridhar, Prashant and Han, Kyu J. and Watanabe, Shinji},
  booktitle={2022 IEEE Spoken Language Technology Workshop (SLT)}, 
  title={E-Branchformer: Branchformer with Enhanced Merging for Speech Recognition}, 
  year={2023},
  volume={},
  number={},
  pages={84-91},
  doi={10.1109/SLT54892.2023.10022656},
  ISSN={},
  month={Jan},}

@inproceedings{NEURIPS2022_9d560961,
 author = {Wei, Jason and Wang, Xuezhi and Schuurmans, Dale and Bosma, Maarten and ichter, brian and Xia, Fei and Chi, Ed and Le, Quoc V and Zhou, Denny},
 booktitle = {Advances in Neural Information Processing Systems},
 editor = {S. Koyejo and S. Mohamed and A. Agarwal and D. Belgrave and K. Cho and A. Oh},
 pages = {24824--24837},
 publisher = {Curran Associates, Inc.},
 title = {Chain-of-Thought Prompting Elicits Reasoning in Large Language Models},
 url = {https://proceedings.neurips.cc/paper_files/paper/2022/file/9d5609613524ecf4f15af0f7b31abca4-Paper-Conference.pdf},
 volume = {35},
 year = {2022}
}

@misc{gulati2020conformerconvolutionaugmentedtransformerspeech,
      title={Conformer: Convolution-augmented Transformer for Speech Recognition}, 
      author={Anmol Gulati and James Qin and Chung-Cheng Chiu and Niki Parmar and Yu Zhang and Jiahui Yu and Wei Han and Shibo Wang and Zhengdong Zhang and Yonghui Wu and Ruoming Pang},
      year={2020},
      eprint={2005.08100},
      archivePrefix={arXiv},
      primaryClass={eess.AS},
      url={https://arxiv.org/abs/2005.08100}, 
}

@InProceedings{pmlr-v202-radford23a,
  title = 	 {Robust Speech Recognition via Large-Scale Weak Supervision},
  author =       {Radford, Alec and Kim, Jong Wook and Xu, Tao and Brockman, Greg and Mcleavey, Christine and Sutskever, Ilya},
  booktitle = 	 {Proceedings of the 40th International Conference on Machine Learning},
  pages = 	 {28492--28518},
  year = 	 {2023},
  editor = 	 {Krause, Andreas and Brunskill, Emma and Cho, Kyunghyun and Engelhardt, Barbara and Sabato, Sivan and Scarlett, Jonathan},
  volume = 	 {202},
  series = 	 {Proceedings of Machine Learning Research},
  month = 	 {23--29 Jul},
  publisher =    {PMLR},
  url = 	 {https://proceedings.mlr.press/v202/radford23a.html}
}

@INPROCEEDINGS{9746312,
  author={Chen, Ke and Du, Xingjian and Zhu, Bilei and Ma, Zejun and Berg-Kirkpatrick, Taylor and Dubnov, Shlomo},
  booktitle={ICASSP 2022 - 2022 IEEE International Conference on Acoustics, Speech and Signal Processing (ICASSP)}, 
  title={HTS-AT: A Hierarchical Token-Semantic Audio Transformer for Sound Classification and Detection}, 
  year={2022},
  volume={},
  number={},
  pages={646-650},
  doi={10.1109/ICASSP43922.2022.9746312}}

@INPROCEEDINGS{10095889,
  author={Elizalde, Benjamin and Deshmukh, Soham and Ismail, Mahmoud Al and Wang, Huaming},
  booktitle={ICASSP 2023 - 2023 IEEE International Conference on Acoustics, Speech and Signal Processing (ICASSP)}, 
  title={CLAP Learning Audio Concepts from Natural Language Supervision}, 
  year={2023},
  volume={},
  number={},
  pages={1-5},
  doi={10.1109/ICASSP49357.2023.10095889}}

@misc{ghosh2025audioflamingo2audiolanguage,
      title={Audio Flamingo 2: An Audio-Language Model with Long-Audio Understanding and Expert Reasoning Abilities}, 
      author={Sreyan Ghosh and Zhifeng Kong and Sonal Kumar and S Sakshi and Jaehyeon Kim and Wei Ping and Rafael Valle and Dinesh Manocha and Bryan Catanzaro},
      year={2025},
      eprint={2503.03983},
      archivePrefix={arXiv},
      primaryClass={cs.SD},
      url={https://arxiv.org/abs/2503.03983}, 
}

@INPROCEEDINGS{10389742,
  author={Gong, Yuan and Liu, Alexander H. and Luo, Hongyin and Karlinsky, Leonid and Glass, James},
  booktitle={2023 IEEE Automatic Speech Recognition and Understanding Workshop (ASRU)}, 
  title={Joint Audio and Speech Understanding}, 
  year={2023},
  volume={},
  number={},
  pages={1-8},
  doi={10.1109/ASRU57964.2023.10389742}}

@misc{ghosh2024gamalargeaudiolanguagemodel,
      title={GAMA: A Large Audio-Language Model with Advanced Audio Understanding and Complex Reasoning Abilities}, 
      author={Sreyan Ghosh and Sonal Kumar and Ashish Seth and Chandra Kiran Reddy Evuru and Utkarsh Tyagi and S Sakshi and Oriol Nieto and Ramani Duraiswami and Dinesh Manocha},
      year={2024},
      eprint={2406.11768},
      archivePrefix={arXiv},
      primaryClass={cs.SD},
      url={https://arxiv.org/abs/2406.11768}, 
}

@InProceedings{10.1007/978-3-031-92611-2_4,
author="Islam, Raisa
and Moushi, Owana Marzia",
editor="Arai, Kohei",
title="GPT-4o: The Cutting-Edge Advancement in Multimodal LLM",
booktitle="Intelligent Computing",
year="2025",
publisher="Springer Nature Switzerland",
address="Cham",
pages="47--60",
isbn="978-3-031-92611-2"
}

@misc{ma2025audiocotexploringchainofthoughtreasoning,
      title={Audio-CoT: Exploring Chain-of-Thought Reasoning in Large Audio Language Model}, 
      author={Ziyang Ma and Zhuo Chen and Yuping Wang and others},
      year={2025},
      eprint={2501.07246},
      archivePrefix={arXiv},
      primaryClass={cs.SD},
      url={https://arxiv.org/abs/2501.07246}, 
}

@inproceedings{zhifei-etal-2025-audio,
    title = "Audio-Reasoner: Improving Reasoning Capability in Large Audio Language Models",
    author = "Xie, Zhifeifinal  and
      Lin, Mingbao  and
      Liu, Zihang  and
      Wu, Pengcheng  and
      Yan, Shuicheng  and
      Miao, Chunyan",
    editor = "Christodoulopoulos, Christos  and
      Chakraborty, Tanmoy  and
      Rose, Carolyn  and
      Peng, Violet",
    booktitle = "Proceedings of the 2025 Conference on Empirical Methods in Natural Language Processing",
    month = nov,
    year = "2025",
    address = "Suzhou, China",
    publisher = "Association for Computational Linguistics",
    url = "https://aclanthology.org/2025.emnlp-main.1216/",
    doi = "10.18653/v1/2025.emnlp-main.1216",
    pages = "23829--23851",
    ISBN = "979-8-89176-332-6"
}

@misc{li2025baichuanomni15technicalreport,
      title={Baichuan-Omni-1.5 Technical Report}, 
      author={Yadong Li and Jun Liu and Tao Zhang and others},
      year={2025},
      eprint={2501.15368},
      archivePrefix={arXiv},
      primaryClass={cs.CL},
      url={https://arxiv.org/abs/2501.15368}, 
}

@misc{xu2025qwen25omnitechnicalreport,
      title={Qwen2.5-Omni Technical Report}, 
      author={Jin Xu and Zhifang Guo and Jinzheng He and Hangrui Hu and Ting He and Shuai Bai and Keqin Chen and Jialin Wang and Yang Fan and Kai Dang and Bin Zhang and Xiong Wang and Yunfei Chu and Junyang Lin},
      year={2025},
      eprint={2503.20215},
      archivePrefix={arXiv},
      primaryClass={cs.CL},
      url={https://arxiv.org/abs/2503.20215}, 
}

@misc{geminiteam2025geminifamilyhighlycapable,
      title={Gemini: A Family of Highly Capable Multimodal Models}, 
      author={Gemini Team and Rohan Anil and Sebastian Borgeaud and Jean-Baptiste Alayrac and others},
      year={2025},
      eprint={2312.11805},
      archivePrefix={arXiv},
      primaryClass={cs.CL},
      url={https://arxiv.org/abs/2312.11805}, 
}

@inproceedings{kim-etal-2019-audiocaps,
    title = "{A}udio{C}aps: Generating Captions for Audios in The Wild",
    author = "Kim, Chris Dongjoo  and
      Kim, Byeongchang  and
      Lee, Hyunmin  and
      Kim, Gunhee",
    editor = "Burstein, Jill  and
      Doran, Christy  and
      Solorio, Thamar",
    booktitle = "Proceedings of the 2019 Conference of the North {A}merican Chapter of the Association for Computational Linguistics: Human Language Technologies, Volume 1 (Long and Short Papers)",
    month = jun,
    year = "2019",
    address = "Minneapolis, Minnesota",
    publisher = "Association for Computational Linguistics",
    url = "https://aclanthology.org/N19-1011/",
    doi = "10.18653/v1/N19-1011",
    pages = "119--132"
}

@INPROCEEDINGS{9052990,
  author={Drossos, Konstantinos and Lipping, Samuel and Virtanen, Tuomas},
  booktitle={ICASSP 2020 - 2020 IEEE International Conference on Acoustics, Speech and Signal Processing (ICASSP)}, 
  title={Clotho: an Audio Captioning Dataset}, 
  year={2020},
  volume={},
  number={},
  pages={736-740},
  doi={10.1109/ICASSP40776.2020.9052990}}

@inproceedings{poria-etal-2019-meld,
    title = "{MELD}: A Multimodal Multi-Party Dataset for Emotion Recognition in Conversations",
    author = "Poria, Soujanya  and
      Hazarika, Devamanyu  and
      Majumder, Navonil  and
      Naik, Gautam  and
      Cambria, Erik  and
      Mihalcea, Rada",
    editor = "Korhonen, Anna  and
      Traum, David  and
      M{\`a}rquez, Llu{\'i}s",
    booktitle = "Proceedings of the 57th Annual Meeting of the Association for Computational Linguistics",
    month = jul,
    year = "2019",
    address = "Florence, Italy",
    publisher = "Association for Computational Linguistics",
    url = "https://aclanthology.org/P19-1050/",
    doi = "10.18653/v1/P19-1050",
    pages = "527--536"
}

@misc{wang2020covost2massivelymultilingual,
      title={CoVoST 2 and Massively Multilingual Speech-to-Text Translation}, 
      author={Changhan Wang and Anne Wu and Juan Pino},
      year={2020},
      eprint={2007.10310},
      archivePrefix={arXiv},
      primaryClass={cs.CL},
      url={https://arxiv.org/abs/2007.10310}, 
}

@inproceedings{melechovsky-etal-2024-mustango,
    title = "Mustango: Toward Controllable Text-to-Music Generation",
    author = "Melechovsky, Jan  and
      Guo, Zixun  and
      Ghosal, Deepanway  and
      Majumder, Navonil  and
      Herremans, Dorien  and
      Poria, Soujanya",
    editor = "Duh, Kevin  and
      Gomez, Helena  and
      Bethard, Steven",
    booktitle = "Proceedings of the 2024 Conference of the North American Chapter of the Association for Computational Linguistics: Human Language Technologies (Volume 1: Long Papers)",
    month = jun,
    year = "2024",
    address = "Mexico City, Mexico",
    publisher = "Association for Computational Linguistics",
    url = "https://aclanthology.org/2024.naacl-long.459/",
    doi = "10.18653/v1/2024.naacl-long.459",
    pages = "8293--8316"
}

@misc{ma2026interspeech2026audioreasoning,
      title={The Interspeech 2026 Audio Reasoning Challenge: Evaluating Reasoning Process Quality for Audio Reasoning Models and Agents}, 
      author={Ziyang Ma and Ruiyang Xu and Yinghao Ma and Chao-Han Huck Yang and Bohan Li and Jaeyeon Kim and Jin Xu and Jinyu Li and Carlos Busso and Kai Yu and Eng Siong Chng and Xie Chen},
      year={2026},
      eprint={2602.14224},
      archivePrefix={arXiv},
      primaryClass={cs.SD},
      url={https://arxiv.org/abs/2602.14224}, 
}

@misc{xu2025qwen3omnitechnicalreport,
      title={Qwen3-Omni Technical Report}, 
      author={Jin Xu and Zhifang Guo and Hangrui Hu and Yunfei Chu and Xiong Wang and Jinzheng He and Yuxuan Wang and Xian Shi and Ting He and Xinfa Zhu and Yuanjun Lv and Yongqi Wang and Dake Guo and He Wang and Linhan Ma and Pei Zhang and Xinyu Zhang and Hongkun Hao and Zishan Guo and Baosong Yang and Bin Zhang and Ziyang Ma and Xipin Wei and Shuai Bai and Keqin Chen and Xuejing Liu and Peng Wang and Mingkun Yang and Dayiheng Liu and Xingzhang Ren and Bo Zheng and Rui Men and Fan Zhou and Bowen Yu and Jianxin Yang and Le Yu and Jingren Zhou and Junyang Lin},
      year={2025},
      eprint={2509.17765},
      archivePrefix={arXiv},
      primaryClass={cs.CL},
      url={https://arxiv.org/abs/2509.17765}, 
}
